\documentclass[11pt,a4paper]{article}
\usepackage{amsmath,amssymb}
\usepackage{epsfig,graphicx}
\usepackage{cite}

\begin{document}

\title{\bf Plasma induced fermion spin-flip conversion $f_L \to f_R + \gamma$}

\author{A.~V.~Kuznetsov\footnote{{\bf e-mail}: avkuzn@uniyar.ac.ru},
N.~V.~Mikheev\footnote{{\bf e-mail}: mikheev@uniyar.ac.ru}
\\
\small{\em Yaroslavl State P.G.~Demidov University} \\
\small{\em Sovietskaya 14, 150000 Yaroslavl, Russian Federation}
}
\date{}

\maketitle

\begin{abstract}
The fermion spin-flip conversion $f_L \to f_R + \gamma$ is considered, 
caused by the difference of the additional energies of the electroweak origin, 
acquired by left- and right-handed fermions (neutrino, electron) in medium. 
An accurate taking account of the fermion and photon dispersion in medium 
is shown to be important. 
\end{abstract}

\section{Introduction}
\label{sec:Introduction}

\def\D{\mathrm{d}} 
\def\E{\mathrm{e}}
\def\I{\mathrm{i}}

The most important event in neutrino physics of the last decades was 
the solving of the Solar neutrino problem. 
The Sun appeared in this case as a natural laboratory for investigations 
of neutrino properties. 
There exists a number of natural laboratories, the supernova explosions, where 
gigantic neutrino fluxes define in fact the process energetics. 
It means that microscopic neutrino characteristics, such as the neutrino 
magnetic moment, etc., would have 
a crucial impact on macroscopic properties of these astrophysical events.

One of the processes caused by the photon interaction with the neutrino 
magnetic moment, which could play an important role in astrophysics, is
the radiative neutrino spin flip transition $\nu_L \to \nu_R \gamma$. 
The process can be kinematically allowed in medium due to its influence 
on the photon dispersion, $\omega = |{\bf k}|/n$ (here $n \ne 1$ is the 
refractive index), when the medium provides the condition $n > 1$. 
In this case the effective photon mass squared is negative, 
$m_\gamma^2 \equiv q^2 < 0$. This corresponds to the well-known 
effect of the neutrino Cherenkov radiation~\cite{Grimus:1993}.  
 
There exists also such a well-known subtle effect as the additional energy $W$ 
acquired by a left-handed neutrino in plasma. 
This additional energy was considered in the 
series of papers by Studenikin et al.~\cite{Grigoriev:2006} as a new 
kinematical possibility to allow the radiative neutrino transition 
$\nu_L \to \nu_R \gamma$. 
The effect was called the ``spin light of neutrino'' ($SL \nu$), and 
later the similar effect ``spin light of electron'' ($SL e$) was discovered. 
For unknown reasons, the photon dispersion in medium providing in part the 
photon effective mass, was ignored in these papers. 
However, it is evident that a kinematical analysis based on the additional 
neutrino or electron energy in matter (caused by the weak forces) 
when the matter influence on the photon dispersion 
(caused by electromagnetic forces) is ignored, cannot be considered as 
a physical approach. Similarly, in the $SL e$ effect 
the authors~\cite{Grigoriev:2006} considered the matter influence on electron 
by the weak forces and ignored the electromagnetic interaction, taking 
the unphysical case of a pure neutron medium. 
It should be noted that even in the conditions of a cold neutron star, 
the fraction of electrons and protons cannot be exacly zero, 
$Y_e \gtrsim 0.01$~\cite{ShapiroTeukolsky}. 
Moreover, even if this unphysical case of a pure neutron medium 
is considered, one should take into account the 
electromagnetic interaction of electrons with the magnetic moments of 
neutrons, which can be much more intensive than the weak 
interaction effects. 

A consistent analysis of the radiative neutrino 
spin flip transition in medium was performed in our 
papers~\cite{Kuznetsov:2006,Kuznetsov:2007}, 
where the medium influence both on the photon and 
neutrino dispersion was taken into account. 
It was shown that the threshold arose in the process, caused 
by the photon (plasmon) effective mass. This threshold left no room for 
the so-called ``spin light of neutrino'' and ``spin light of electron'' 
in the real astrophysical situations. 

In the series of papers~\cite{Kouzakov:2008} the authors declare that they 
have developed a powerful method of exact solutions of the modified Dirac 
equations including the effective matter potentials.

In this paper, we remind the basic points of our criticism upon the $SL \nu$ 
effect, and give some comments on the method of exact 
solutions~\cite{Kouzakov:2008}.

\section{Additional left-handed neutrino energy and effective mass}
\label{sec:additional}

As it was already mentioned, the effect of the ``spin light of neutrino'' 
proposed in~\cite{Grigoriev:2006} was based on the additional 
left-handed neutrino energy $W$ induced by the medium influence. 
Just this additional energy  
provides an effective mass squared $m_L^2$ to the left-handed neutrino, 
\begin{equation}
m_L^2 = {\cal P}^2 = (E + W)^2 - {\bf p}^2 = 2 \, E \, 
W + W^2 + m_\nu^2\,, 
\label{eq:effective_mass}
\end{equation}
where ${\cal P}$ is the neutrino four-momentum in medium, while 
$(E,\, {\bf p})$ would form the neutrino four-momentum in vacuum, 
$E = \sqrt{{\bf p}^2 + m_\nu^2}$. It will be further seen that 
the neutrino vacuum mass $m_\nu$, a great attention was paid to in the $SL \nu$ 
analysis~\cite{Grigoriev:2006}, may safely be neglected. 

Given a $\nu_L \nu_R \gamma$ interaction, caused by the neutrino magnetic moment, 
the left-handed neutrino effective mass $m_L$~(\ref{eq:effective_mass}) would open 
a kinematical possibility for the process $\nu_L \to \nu_R + \gamma$,
if the photon effective mass is less than $m_L$. 

Basing the consideration of the ``spin light of neutrino'' 
on the additional left-handed neutrino energy, 
the authors~\cite{Grigoriev:2006}, nevertheless, did not analyse 
this value in detail. 
  
The expression for this additional energy of 
a left-handed neutrino with the flavor $i = e, \mu, \tau$ was obtained in the 
local limit of the weak interaction~\cite{Notzold:1988,Pal:1989,Nieves:1989}, 
see also Ref.~\cite{Elmfors:1996}, and can be presented in the following 
form
\begin{eqnarray}
&&W_i = \sqrt{2} \, G_{\rm F} \left[
\left(\delta_{ie} - \frac{1}{2} + 2 \, \sin^2 \theta_{\rm W}\right) 
\left(N_e - \bar N_e \right) + \left(\frac{1}{2} 
- 2 \, \sin^2 \theta_{\rm W}\right) \left(N_p - \bar N_p \right) \right.
\nonumber\\
&& 
- \left. \frac{1}{2} \left(N_n - \bar N_n \right) 
+ \sum\limits_{\ell = e, \mu, \tau} \left(1 + \delta_{i \ell} \right)
\left(N_{\nu_\ell} - \bar N_{\nu_\ell} \right) 
\right] , 
\label{eq:EnuLgen}
\end{eqnarray}
where the functions 
$N_e, N_p, N_n, N_{\nu_\ell}$ are the number densities of background electrons, 
protons, neutrons, and neutrinos, and $\bar N_e, \bar N_p, \bar N_n, 
\bar N_{\nu_\ell}$ are the densities of the corresponding antiparticles.
To find the additional energy for antineutrinos, one should change the 
total sign in the right-hand side of Eq.~(\ref{eq:EnuLgen}). 

As is seen from Eq.~(\ref{eq:EnuLgen}), this value becomes zero in the 
charge-symmetric plasma. This means that the local 
limit of the weak interaction does not describe comprehensively the 
additional neutrino energy in plasma, and the non-local weak contribution 
must be taken into account. The analysis of this contribution was first 
performed for the conditions of the early 
Universe~\cite{Notzold:1988,Elmfors:1996}. 

The non-local weak contribution into the additional 
neutrino energy in plasma, which is identical for both neutrinos and 
antineutrinos, can be presented in the form
\begin{equation}
\Delta^{(\rm{nloc})} W_i = 
- \frac{16 \, G_{\rm F} \, E}{3 \, \sqrt{2}} 
\left[ \frac{<E_{\nu_i}>}{m_Z^2} 
\left(N_{\nu_i} + \bar N_{\nu_i} \right)  
+ \delta_{ie} \, \frac{<E_e>}{m_W^2} \left(N_e + \bar N_e \right)
\right] , 
\label{eq:W_nloc}
\end{equation}
where $E$ is the energy of a neutrino propagating through plasma, $<E_{\nu_i}>$ 
and $<E_e>$ are the averaged energies of plasma neutrinos and electrons 
correspondingly. In a particular case of a charge symmetric hot plasma, 
this expressions reproduces the result of 
Refs.~\cite{Notzold:1988,Elmfors:1996}:
\begin{equation}
\Delta^{(\rm{nloc})} W_i = 
- \frac{7 \, \sqrt{2} \, \pi^2 \, G_{\rm F} \, T^4}{45} 
\left( \frac{1}{m_Z^2} + \frac{2 \, \delta_{ie}}{m_W^2} \right) E \, .
\label{eq:W_early}
\end{equation}
The minus sign in~(\ref{eq:W_early}) 
unambiguously shows that in the early Universe the process of the radiative 
spin-flip transition is forbidden both for neutrinos and antineutrinos. 

The absolute value of the non-local weak contribution~(\ref{eq:W_nloc}) 
grows with the neutrino energy. It means that this contribution can 
be essential at ultra-high neutrino energies.

\section{Does the window for the ``spin light of neutrino'' exist?}
\label{sec:window}

\begin{figure}[htb]
\centering
\includegraphics[width=0.95\textwidth]{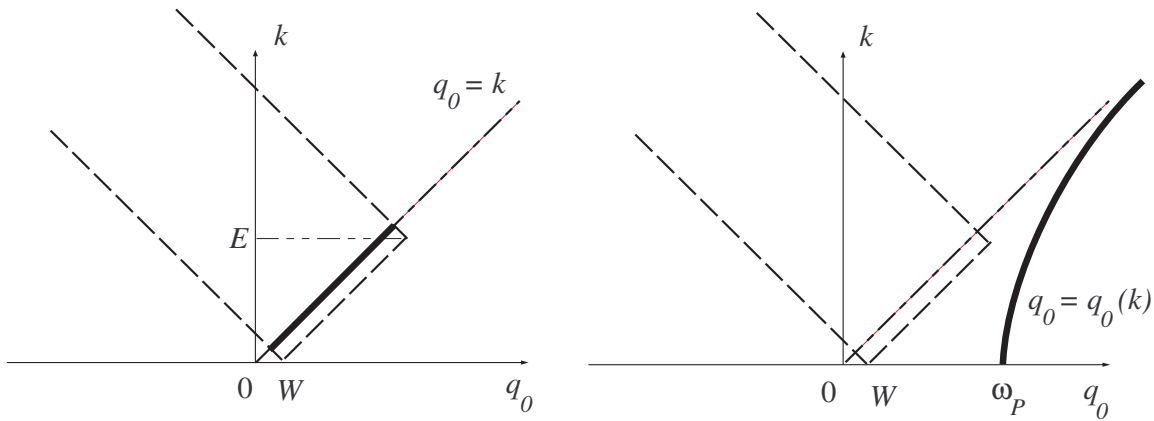}
\caption{
The region of integration for the $\nu_L \to \nu_R$ conversion width 
with the fixed initial neutrino energy $E$ is inside the slanted rectangle 
shown by dashed line. The vacuum photon dispersion (if the medium influence 
is ignored) is shown by bold line in the left plot. 
The photon dispersion curve in plasma is shown by bold line in the right plot.
}
\label{fig:disp}
\end{figure}

To show manifestly that the case considered in the papers by 
Studenikin et al.~\cite{Grigoriev:2006}, with taking the additional left-handed 
neutrino energy $W$ in plasma and ignoring 
the photon dispersion, was really unphysical, let us 
consider the region of integration for the 
$\nu_L \to \nu_R$ conversion width.   
In Fig.~\ref{fig:disp}, the photon vacuum dispersion line $q_0 = k$ 
is inside the allowed kinematical region (left plot), but 
the plasma influenced photon dispersion curve appears to be outside, 
if the neutrino energy is not large enough (right plot). 
 
For the fixed plasma parameters, the threshold neutrino energy 
$E_{\mathrm{min}}$ exists for coming of the dispersion curve into 
the allowed kinematical region. Even for the interior of a 
neutron star this threshold neutrino energy is rather large: 
\begin{equation}
E_{\mathrm{min}} \simeq \frac{\omega_P^2}{2 \, W} 
\simeq 10 \,{\rm TeV} \,, 
\label{eq:threshold}
\end{equation}
where $\omega_P$ is the plasmon frequency.

One could hope that the ``spin light of neutrino'' may be possible at 
ultra-high neutrino energies. However, in this case the local limit of 
the weak interaction is incomplete, and the non-local weak 
contribution into additional neutrino energy $W$ 
must be taken into account. This contribution 
has always a negative sign, and its absolute value grows with the 
neutrino energy. One could only hope that 
the window arises in the 
neutrino energies for the process to be kinematically opened, 
$E_{\mathrm{min}} < E < E_{\mathrm{max}}$. 
For example, in the solar interior there is no window 
for the process with electron neutrinos at all. 
A more detailed analysis of this subject was performed in our 
papers~\cite{Kuznetsov:2006,Kuznetsov:2007}. 

\section{``Exact solutions'' of inexact equations} 

To construct the amplitudes of the fermion spin-flip conversion 
processes $f_L \to f_R + \gamma$, it is necessary to know the solutions 
of the Dirac equation including the matter effects. 
In the series of papers~\cite{Kouzakov:2008} the authors declare that they 
have developed a powerful method of exact solutions of the modified Dirac 
equations in matter. 

First of all, there is some progress in the last papers~\cite{Kouzakov:2008} 
with respect to the $SL \nu$ effect. 
Namely, the five completely unphysical regions of parameters, 
which were discussed in details in the previous papers, are removed from the analysis, 
and only one case of an ultra-high neutrino energy is considered.
However, an essential threshold effect is still not mentioned at all. 
An incorrect statement is repeated in~\cite{Kouzakov:2008} 
that our results~\cite{Kuznetsov:2006,Kuznetsov:2007} for the case of an ultra-high 
neutrino energy exactly reproduce the results of the authors~\cite{Grigoriev:2006}. 
In fact, the width for the process $\nu_L \to \nu_R + \gamma$ obtained in our 
papers~\cite{Kuznetsov:2006,Kuznetsov:2007} had much more general form being valid 
for arbitrary neutrino energies above the threshold, 
while the process width presented in~\cite{Grigoriev:2006} could be valid 
for the neutrino energies much greater than the threshold. 

As in the previous papers, 
the non-local weak contribution into the additional neutrino energy in plasma 
is not taken into account in~\cite{Kouzakov:2008}, 
while it is essential at ultra-high neutrino energies.
Without this non-local weak contribution, the Dirac equation in medium 
for a neutrino is approximate by definition, and the term of an ``exact 
solution'' becomes dubious. 

Making an attempt of constructing a new approach to the description of the neutrino 
and electron processes in matter, the authors~\cite{Kouzakov:2008} refer to the 
method of exact solutions developed for the processes in a strong external 
electromagnetic field. However, it is not a good justification. 
The strong field influence on the properties of charged particles 
is the essential non-perturbative effect where the analysis of the quantum processes, 
based just on exact solutions, is required. On the other hand, the matter 
influence on the neutrino and electron processes due to the weak interaction is essentially 
perturbative in any conceivable astrophysical conditions. 

Moreover, the explicit form of the modified Dirac equation 
is rather simple in the low-energy approximation only, 
when the modification is caused by the additional 
left-handed neutrino energy in the form~(\ref{eq:EnuLgen}) calculated in the 
local limit of the weak interaction. As it was mentioned above, for 
high neutrino energies the non-local weak contribution~(\ref{eq:W_nloc}) 
growing linearly with the neutrino energy, appears to be essential. 
And even this non-local term~(\ref{eq:W_nloc}) is nothing but the result  
of the expansion of the $W$-boson propagator over the parameter $Q^2/m_W^2$. 
It is obvious that such an expansion has a physical sence for the neutrino energies 
$E \ll m_W^2/m_e \sim 10^4\ $ TeV. 
So, if one pretends to describe the additional left-handed neutrino energy 
in matter for the neutrino energies much higher the threshold 
energy~(\ref{eq:threshold}), an exact calculation is required without 
expanding the $W$-boson propagator. In this case the modified Dirac 
equation should take the form of the integro-differential equation. 
Without such an analysis, speculations on exact solutions of the modified Dirac 
equation in matter have no ground. 
 
\section{Conclusion}

\begin{itemize}
\item
We have shown that an approach based on a subtle effect of the medium influence 
on the neutrino dispersion, when the much more significant influence of the same 
medium on the photon dispersion is ignored, has no physical sense.
\item
With the photon dispersion taken into account, the threshold 
neutrino energy exists for the process $\nu_L \to \nu_R + \gamma$, which is very large. 
\item
At ultra-high neutrino energies, the non-local weak contribution 
into the additional neutrino energy in plasma must be taken into account. 
There arises the window (if exists) in the 
neutrino energies for the process to be kinematically opened, 
$E_{\rm{min}} < E < E_{\rm{max}}$. 
\item 
Without the non-local weak contribution into the additional neutrino 
energy in plasma, the Dirac equation in medium 
for a neutrino is approximate by definition, and the term of an ``exact 
solution'' becomes dubious. 
\end{itemize}

\section*{Acknowledgments}

We thank S.I. Blinnikov, M.I. Vysotsky, V.A. Novikov, L.B. Okun for useful 
discussion. We express our deep gratitude to the organizers of the 
Seminar ``Quarks-2008'' for warm hospitality.

The work was supported in part 
by the Council on Grants by the President
of the Russian Federation for the Support of Young Russian Scientists
and Leading Scientific Schools of Russian Federation under the Grant 
No.~NSh-497.2008.2, and 
by the Russian Foundation for Basic Research under the Grant No.~07-02-00285-a.



\end{document}